
\magnification=\magstep1
\hsize=15truecm
\overfullrule=0pt
\vsize=20truecm
\hoffset=1truecm
\voffset=1truecm
\baselineskip=24pt
\hfill 11/11/94
\centerline{GAUSSIAN FLUCTUATION IN RANDOM MATRICES\footnote{*}{Work
supported in part by NSF Grant DMR 92-13424.}}
\bigskip
\centerline{by}
\bigskip
\centerline{Ovidiu Costin and Joel L. Lebowitz}
\centerline{Department of Mathematics and Physics}
\centerline{Rutgers University, New Brunswick, New Jersey 08903}
\bigskip
\bigskip
\noindent {\bf Abstract}
Let $N(L)$ be the  number of eigenvalues,
in an interval of length $L$, of a
matrix chosen at random from the
Gaussian Orthogonal, Unitary or Symplectic ensembles of ${\cal N}$ by ${\cal
N}$ matrices, in the limit  ${\cal N}\rightarrow\infty$.
We prove that $[N(L) - \langle N(L)\rangle]/\sqrt{\log L}$ has a Gaussian
distribution when $L\rightarrow\infty$.
This theorem, which
requires control of all the higher moments of the distribution,
elucidates numerical and exact results on chaotic quantum systems and
on the statistics of zeros of the Riemann
zeta function.
\bigskip
\noindent PACS nos. 05.45.+b, 03.65.-w
\eject

Ensembles of ${\cal N}$ dimensional Gaussian Random Matrices (GRM) with
invariances under the Orthogonal, Unitary or Symplectic groups
corresponding to the GOE, GUE and GSE were introduced by Wigner and
developed by Porter, Dyson, Mehta and others [1,2].  Wigner's inspired
surmise that the statistics of eigenvalues of these GRM can be used to model
the statistical properties of the observed spectra of complex nuclei turned
out to be exactly right.  There is indeed good agreement between the
observed high energy level spacings, pair correlations and variance or
$\Delta$-statistics and those calculated analytically from the GRM in the
limit ${\cal N} \to \infty$.  Moreover, the GRM have been found to be the
very robust ``renormalization group fixed points'' of a large class of RM
[3] which play an important role in many areas of physics and mathematics
[1--7].

In the present work we focus on the large $L$ (long wavelength) behavior of
the random variable $N(L)$ giving the number of eigenvalues of a GRM,
chosen from any of the Gaussian ensembles, in an interval $(y, y+L)$:  We
always consider the ${\cal N} \to \infty$ limit when the distribution is
translation invariant and use units in which the mean spacing is unity.
It is well known that the variance of $N(L)$ grows like $\log L$ as $L \to
\infty$.  We prove that {\it all} the moments of $\xi(L) \equiv
[N(L)-L]/\sqrt{\log L}$ approach for large $L$ those of a Gaussian
distribution which implies (weak) convergence of $\xi(L)$ to a Gaussian
random variable.  We shall discuss later the connection of our result with
the statistics of energy levels of quantum systems with generic chaotic
classical Hamiltonians and of the zeros of the Riemann zeta function [4,5,7].

It is a remarkable fact that the distribution
of energy levels of the G(O,U,S)E
are given by the Gibbs canonical distribution of
the positions of  charged point particles on the line interacting via the (two
dimensional) repulsive logarithmic Coulomb potential, $v(r) = -\log r$,
at  reciprocal temperatures $\beta = 1, 2, 4$ respectively [1,2].
The particles with  positions $x_i$, $i = 1, ..., \cal N$,  on the real line,
are confined  by a uniform negative background, which produces a
harmonic potential.  The total potential energy of the system is
$$
V_G(x_1, ..., x_{\cal N}) = {1 \over 2} \sum_{i=1}^{\cal N} x_i^2 - {1
\over 2} \sum_{i\ne j}^{\cal N} \log |x_i - x_j|.\eqno(1)
$$
In the corresponding circular ensembles of Dyson
the $x_i$ lie on a circle of length $\cal N$ and the energy is given
by the second term on the r.h.s of (1)
with  distance measured in the plane.
The canonical Gibbs measures corresponding to Gaussian and circular
ensembles
become equivalent in the thermodynamic limit, $\cal
N \to \infty$, yielding the same $k$-point, $k=1,2,3,..$
correlation functions, for all $\beta > 0$ [1,2,8].

These infinite volume correlation functions are known explicitly for the
``solvable'' cases $\beta = 1, 2, 4$ corresponding to the GRM.
Defining as usual $n_j(x_1,...,x_j)$ as the joint density
for $j$-tuples, the corresponding Ursell functions \break $U_k(x_1,...,x_k)$
[9] are given by
$$
U_1(x_1) \equiv n_1(x_1) \equiv 1\eqno(2)
$$
$$
U_2(x_1,x_2) \equiv n_2(x_1,x_2) - n_1(x_1)n_1(x_2) = n_{2}(x_1-x_2)-1
$$

$$
U_k(x_1,...,x_k) \equiv \sum_G (-1)^{m-1} (m-1)! \prod_{j=1}^m
n_{G_j}(\{x_{G_j}\})
$$
where $G$ is a partition of the indices $(1,2,..,k)$ into $m$ subgroups
$G_1,..,G_m$ and $\{x_{G_j}\}$ are the $x_i$ with indices in $G_j$.
(In the GRM literature,  $Y_k=(-1)^{k-1}U_k$ is usually called the
$k^{\rm th}$ cluster function). The integrals,
$\bar U_k$, of $U_k$ over
a k-dimensional cube having sides of length L are  directly related to the
cumulants $C_j(L), j=1,...,k$  of the random variable $N(L)$, the number of
points (or
eigenvalues) in an interval of length $L$, which we shall take for
 definiteness  to be the interval $[-t,t]$, $L\equiv 2t$.  Thus
$$
\bar U_1 =\int_{-t}^t U_1(x_1)dx_1 = 2t = \langle N(L) \rangle =
C_1(L)\eqno(3)
$$
$$
\eqalign {\bar U_2 = \int_{-t}^t \int_{-t}^t dx_1 dx_2 U_2(x_1,x_2) &\equiv
\langle N(N-1)
\rangle - \langle N \rangle^2 = \langle (N - \langle N \rangle)^2 \rangle -
\langle N \rangle\cr  &= C_2(L) - C_1(L)}
$$
$$
\eqalign {\bar U_3 &= C_3(L) - 3C_2(L) + 2C_1(L)},\ \hbox{etc}
$$

Using the generating function [9]
$$
F(\mu) = \sum_{n=0}^{\infty} E_n(L) e^{n\mu},\eqno(4)
$$
with $E_{n}(L)$ the probability of having exactly $n$ particles in the
interval $L$, we  have,
$$
\log F(\mu) = \sum_{n=1}^{\infty}{ C_n\over n!} \mu^n =
\sum_{n=1}^{\infty}{1
 \over n!} \bar U_n(e^{\mu}-1)^n.\eqno(5)
$$
This gives

$$
C_k = \left(\sum_{j=2}^{k-1} b_{k,j} C_j\right) + (-1)^k (k-1)!C_1
+ \bar U_k \eqno(6)
$$
where
$$
\eqalign {b_{n,j}&= b_{n-1,j-1} - (n-1)b_{n-1,j},\ \ \ \ \ 2 \leq j\leq n-1\cr
b_{n,n} &= -1\hskip 120pt n\ge 2}
$$

The $U_k$
take on a particularly simple form for the GUE ($\beta=2$) [1,2],
$$
U_k(x_1,...,x_k) = (-1)^{k-1} \sum_{\hbox{Perm}} \prod_{i=1}^k S(x_{i+1} - x_i)
\eqno(7)$$
where $S(x) =(\pi x)^{-1}\sin \pi x$ and $x_{k+1}=x_1$, so the indices are to
be
thought of as being on a circle.  Using (6) one readily obtains [1,2]
$$
C_2(L) = (\log L)/\pi^2 + O(1).\eqno(8)
$$

Note that for a system with short range interactions $C_2(L)$
would
grow like $L$ but the logarithmic interactions
between the (charged) particles induce a sort of
local crystalline order reducing the variance to $\log L$. It is this
strong correlation which produces level repulsion between the
eigenvalues and makes the large scale behavior
of the fluctuations far from obvious.
Defining now the normalized random variable
$$
\eta(L) = \left(N(L) - L\right)/\sqrt {{1 \over \pi^2} \log L}\eqno(9)
$$
the $k^{\rm th}$ cumulant of $\eta(L)$ will be $c_k\equiv
C_{k}/[\log L/\pi^2]^{k \over 2}$.
As is well known c.f.\ [10], $\eta(L)$ will approach a Gaussian
random variable (with mean zero and unit variance) as $L \to \infty$,
if and only if  all $c_k$, $k \geq
3$, go to zero,  i.e. if
$C_k=o([\log L/\pi^2]^{k \over 2})$. Using an induction argument based on the
recurrence
relation
(6) and the equality (7), this
corresponds to proving that
$$
\eqalign{
s_{k}(t)&\equiv\int_{-t}^t ... \int_{-t}^t dx_1..dx_k S(x_2-x_1)
S(x_3-x_2)...S(x_1-x_k)
\cr
&= 2t+o[(\log t)^{k\over 2}], k \geq 3}\eqno(10)
$$

\noindent We shall actually prove that $s_k(t) = 2t+O(\log t)$ which implies
that, for $k \geq 3$ $C_k(L)=O(\log L)$; in fact we believe
that for $k \geq 3$, $C_{k}(L)$ stays bounded as $L \to \infty$,
as suggested by the explicit asymptotic evaluation
of the integrals.

$$\eqalign{
s_3& = 2t-{3\over{2\pi^2}}\log t+O(1)\cr
s_4&= 2t-{11\over{6\pi^2}}\log t+O(1)}\eqno(11)
$$
which gives, using (6), that $C_3$ and $C_4$ are of $O(1)$.

\bigskip

\noindent To prove (10) we make use of the fact [1] that
$$
s_k(t) = \sum_{i=1} \lambda_i^k (t) = {\rm Tr}\, {\bf S}^k(t)\eqno(12)
$$
where the $\lambda_i (t)$ are the eigenvalues of the integral operator
\break ${\bf S}(t), ({\bf S}f)(x) = \int_{-t}^t dy S(x-y) f(y)$.
It is known [1] (and can be easily proven) that the spectrum of ${\bf S}$
lies in $[0,1]$.  We can now use an induction argument to prove that
$s_{k}(t) = 2t + O(\log t)$.  This is so for $k = 1, 2$ (also for $3,
4$), and for $k \geq 2$ we have

$$ {\rm Tr\,}{\bf S}^{k+1} = {\rm Tr\,}{\bf S^k}
 - {\rm Tr}\,\big[{\bf S}^{k-1}({\bf S}- {\bf S}^{2})
\big].\eqno(13) $$

\noindent  The first term is of the desired form by the
induction assumption while the terms in the parenthesis are positive
operators, and $\|{\bf S}^{k-1} \| \leq 1$ so it can be taken out of the
product yielding $ {\rm Tr\,}{\bf S}^{k+1} = {\rm Tr\,}{\bf
S^k}+O(\log t)$
and the proof is complete.

Our results readily extend to show that if we divide up the real line into a
union of
intervals of length $L$, let $N_j(L)$ be the particle
number in $[jL, (j+1)L]$ and set $
\eta_j(L) = [N_{j}(L) - L]/\sqrt{C_2(L)},\ j\in Z
$
then the $\{\eta_j(L)\}$ approach, as $L \to \infty$,  jointly Gaussian
random variables with mean zero and covariances
$
\langle \eta_j \eta_k \rangle = \delta_{j,k} - {1 \over 2} \delta_{j \pm 1,
k}.
$

To prove the results for the GOE, $\beta = 1$, we use an identity
conjectured by Dyson and proved by Gunson [11] (we thank Freeman Dyson
for pointing this out to us).  According to this identity,
superimposing two noninteracting Coulomb gases, say blue and
red, in the circular ensemble at reciprocal
temperature $\beta = 1$ and then looking only at alternate particles,
e.g. at all the odd (or even) ones, yields the distribution
at $\beta=2$. Considering now the
number of
particles in an interval of length $L$ gives
$$ N_{total} (L) =
N_{blue}^{(1)} (L) + N_{red}^{(1)} (L) = N_{odd}^{(2)} (L) +
N_{even}^{(2)}
 (L) =
2N^{(2)} (L) + (0, \pm 1)\eqno(14) $$

\noindent where the superscripts $(1,2)$
stand for the random variables obtained from the ensembles with
$\beta = 1, 2$, and $N_{blue}, N_{red}$ are independent.  This shows
immediately that in the infinite $\cal N$ limit, the variables
$N^{(2)} (L)$ and $N^{(1)} (L)$ normalized by the square root of their
variances have the same asymptotic behavior.  Taking $\langle
N_{blue}^{(1)} (L) \rangle = \langle N_{red} ^{(1)} (L) \rangle = \langle
N^{(1)} (L) \rangle = L$ we have

$$\eqalign{
{1\over 2}\big
\langle \big[(N_{blue}^{(1)} (L) - L)
 + (N_{red}^{(1)} (L) - L)\big]^2 \big\rangle &=  \langle
(N^{(1)} (L) - L)^2 \rangle = \cr
&2\langle (N^{(2)} (L) - L)^2 \rangle
\sim{{2\log L}\over{\pi^2}}}\eqno(15)
$$
giving the well known variance of the GOE [1].

For the GSE, $\beta = 4$, we use the equality between the statistics of its
eigenvalues and the odd eigenvalues of the GOE [1].  This again leads
to Gaussian asymptotics with a
variance given by ${1\over{2\pi^2}}\log L+O(1)$.
It seems very reasonable to expect and one can give strong heuristic
arguments, based on the ``long wavelength response'' of ``Coulomb'' systems,
that the Gaussian
nature of the fluctuations, with variances $(2/\pi^2\beta)\log L$,
 holds for all $\beta$. ( We
thank Bernard Jancovici for pointing this out to us, see also [8].)

Using more detailed information on the spectrum of ${\bf S}$ (see
[12]), it follows that $s_k(t)=2t-\pi^{-2}\sum_{j=1}^{k-1}j^{-1}\,\log
t+o(\log t)$ which using (6) implies that $C_k=o(\log t)$ for
all $k\ge 3$. (We are indebted to Harold Widom for this
information). Widom also noted that
our proof does not make any use of the specific form of
${\bf S}$. It only uses the
property spec$(S)\in[0,1]$ and the fact that $Tr({\bf S}-{\bf
S}^2)\rightarrow\infty$ (as $t\rightarrow\infty$). The conclusion
therefore holds
for a larger class of  matrix models [3].

The local statistics of the eigenvalues $\epsilon_j,
j=1,2,...,\infty$, $0\le\epsilon_1\le\epsilon_2\le...$ of a
classically chaotic quantum Hamiltonian (CQH), such as the geodesic flow
on a (non-arithmetic) surface of constant negative curvature or the
Sinai billiard appear to coincide at high energies with
those obtained from the GRM [4]. More precisely,
if we consider the energy levels
of a generic CQH, suitably scaled so that the mean distance
between levels is unity, in an interval $(y, y+L)$ then their
statistics, obtained by letting $y$ vary uniformly in some
interval
$(T_1,T_2)$ will coincide, for $T_2\rightarrow\infty$, with
that obtained from one of the standard GRM ensembles when the matrix
size tends to infinity. Our result then predicts a Gaussian
distribution
of the fluctuations in the number of levels $N(L)$ when
$L\rightarrow\infty$.
A numerical check of this for some CQH will require using energy
levels in a (scaled) energy interval $L$ with $1<\!\!< L<\!\!< (T_2-T_1)$
, $T_1$ large.

It is
also interesting , as emphasized by Berry [4], to consider in addition to
the local statistics of quantum levels also their
global statistics. These correspond, in our context, to
fluctuations in the number of levels in an
interval $(y,x)$ whose length $L(x)=x-y(x)$ is not fixed
but grows with $x$ as $x$ varies in the interval $(T_1,T_2)$
with $T_2\rightarrow\infty$. This
includes in particular the case $y=0,\ L(x)=x$ corresponding
to the fluctuations in the number of eigenvalues less than $x$. This
quantity, normalized by the square
root of its variance, was conjectured in [5] (where it is denoted by
$N_{fl}(x)$) to have a Gaussian
distribution as $x\rightarrow\infty$ for all CQH.  If true this would be a
general characterization of CQH and distinguish them from integrable systems
where
it was found rigorously that
the global distribution is non-Gaussian [13]. Quite generally, it was shown
by Berry [5] that when $L(x)>L_{\max}(x)$,
the variance of $N(L(x))$
saturates for $L>L_{\max}$. Berry also found that $N_{fl}^2(x)$
(averaged over some interval containing many eigenvalues
but very small compared to $x$) grows for billiard systems
like
$(2\pi)^{-2}\log x$.

As already noted,
the distribution of eigenvalues in the GRM  is translation
invariant, when the matrix size ${\cal N}$ goes to infinity,
so there is no analog of $L_{\max}$ in our considerations.
One can
however
consider fluctuations in $N(L)$ for an interval $L(\cal N)$ which contains
a number of eigenvalues small compared to $\cal N$ but goes to infinity
when $\cal N \to \infty$, e.g.\ in the circular
ensemble we could have $L(N) \sim N^{\gamma}, \gamma < 1$ or even like
$\alpha {\cal N}, \alpha << 1$. For the Coulomb
system with neutralizing background it is also possible to consider
semi-infinite systems with various boundary conditions and/or nonuniform
background. Some such systems have been considered
in [8] and
we believe that our results about
Gaussian behavior
would  extend also to these systems which might model
some of the saturation
features of CQH.

We turn finally to the (non trivial)
 zeros of the Riemann zeta function $\zeta(z) =
\sum_{n=1}^{\infty} n^{-z}$  which are, according to the Riemann Hypothesis,
of the form \hbox{$z_n={1\over 2}+i\gamma_n$}. As pointed out by Berry [4]
there are reasons to expect similarities between the statistics of the
$\gamma_n$ and of energy levels of CQH.  In fact Montgomery [14] proved
that the pair correlation function of the $\gamma_n$ agrees with that of
the GUE, Eq. (7).  Numerical calculations by
Odlyzko [7] give striking evidence that the nearest neighbor level
spacing distribution of the $\gamma_n$ is, for large $n$, indeed the
same as that obtained from the GUE.  In a very interesting recent paper
Rudnick and Sarnak [7] greatly extended the results of Montgomery by
showing that the $n$-point correlation functions of these zeros converge,
on a large class of test functions to those of the GUE.  Moreover the
normalized global
fluctuation in these zeros corresponding to $N_{fl}(x)$, was shown by
Selberg [15]  to have a
Gaussian distribution. The same arguments imply that the local fluctuation in
their
number in an interval  $(y,y+L)$ averaged over $(T_1,T_2)$ and properly
scaled  becomes Gaussian when $T_2 \rightarrow\infty$,
followed by $L\rightarrow\infty$ (we are indebted to Peter Sarnak for
explaining this to us).  The results proved here thus fit completely
with the picture of the statistics of Riemann zeros being in the same
``universality class'' as that of the GUE.

\bigskip
\vskip -1em
\noindent {\bf Acknowledgments}

We have benefited much from discussions with F. Dyson, B. Jancovici, M.L.
Mehta and P. Sarnak.  We also thank M.V. Berry, E. Brezin,
H.Widom for very  useful comments and  J. Bolte and F. Steiner for calling
our attention to their work prior to publication.
\bigskip
\vskip -1em
\noindent {\bf References}
\item {[1]} \ For a collection of original articles and an overview see
$\underline{\hbox{Statistical Theories}}$
$\underline{\hbox{ of Spectra: Fluctuation}}$,
C.E. Porter, ed., Academic Press (1965).
M.L. Mehta, $\underline{\hbox{Random Matrices}}$, 2nd edition, Academic Press
(1990).

\item {[2]}  \ F.J. Dyson, J.Math. Phys. $\underline{3}$, 166, 1191,
1199
(1962); Commun. Math. Phys. $\underline{19}$, 235 (1970),
$\underline{47}$, 171 (1976).

\item {[3]} \ E. Brezin and A. Zee, Nucl. Phys. B402 [FS] 613 (1993).
L.Pastur, Lett. Math. Phys ${\underline{25}}$, 259 (1992) and preprint. See
also H. Leff, J. Math. Phys. $\underline{5}$, 756, 763 (1964) and D. Fox
and P.B. Kahn, Phys. Rev. $\underline{134}$, B1151 (1964).

\item {[4]} \ M. V. Berry, Proc. Roy. Soc. London A ${\underline{400}}$,229
(1985); $\underline {{\rm Nonlinearity} 1}$, 399 (1988);  see also articles by
Berry and others in
 Les Houches Lectures L II, 125,
M.J. Giannoni, A. Voros and J. Zinn-Justin eds. North Holland (1991).
O. Bohigas, M.J. Giannoni and C.S. Schmit
in
$\underline{\hbox{Quantum Chaos and}}$ $\underline{\hbox{Statistical
 Nuclear Physics}}$, T. H. Zelegman and H.
Nishioka, eds.
LNP $\underline{263}$, Springer (1986).
M. C. Gutzwiller, $\underline{\hbox{Chaos in Classical and Quantum
Mechanics}}$, Springer (1990).

\item {[5]} \  R. Aurich, J. Bolte and F.Steiner, Phys. Rev. Lett
${\underline{73}}$, 1356 (1994).

\item {[6]} \  M. Moshe, H. Neuberger and B. Shapiro, Phys. Rev. Lett
${\underline{73}}$, 1497 (1994).
E. Basor, C.A. Tracy and H. Widom,  Phys. Rev. Lett
${\underline{69}}$, 5 (1992).
G. Mahoux and M. L. Mehta, J. Phys. J. France  ${\underline{3}}$,
697 (1993).

\item {[7]} \ A.M. Odlyzko, AT\&T preprint (1989). Z. Rudnick and P. Sarnak,
Princeton preprint (1994).

\item {[8]} \ P. J. Forester, Nucl. Phys. B $\underline{402}$, 709 and
$\underline{416}$, 377 (1994).
B. Jancovici and P.J. Forester, Phys. Rev. B (1994) and  preprints.

\item {[9]} \ J.L. Lebowitz and J.K. Percus,  J. Math Phys.
$\underline{4}$, 248 (1963).

\item {[10]} \ c.f. L. Breiman, $\underline{\hbox{Probability}}$,
Addison-Wesley (1968).

\item {[11]} \ J. Gunson, J. Math. Phys $\underline{1}$, 4 752
 (1962).

\item {[12]} \  H.J. Landau and H. Widom, J. Math. Anal. and Appl. 77 469-481
(1980).

\item {[13]} \  P. Bleher, F. Dyson, J.L. Lebowitz, Phys. Rev. Lett.
$\underline{71}$, 3047 (1993).

\item {[14]} \  H.L. Montgomery, Proc Symp. Pure Math $\underline{24}$,
181,
Am. Math. Soc., Providence (1973).

\item {[15]}  \ A. Selberg, Arch. Math. Naturvid. $\underline{B48}$, 89
(1946):  $\underline{\hbox{Collected Papers}}$, Vol. II, pp. 47 (Springer,
1991).

\end